\documentclass[showpacs,amssymb,floatfix]{revtex4}[14pt]
\usepackage{graphicx}
\usepackage{dcolumn}
\usepackage{graphics}
\usepackage{epstopdf}
\usepackage{pdfpages}
\usepackage{bm}
\begin{document}
\title{Bound states in the continuum with high orbital angular momentum
in a dielectric rod with periodically  modulated permittivity}
\author{Evgeny N. Bulgakov$^{1,2}$ and Almas F. Sadreev$^1$}
\affiliation{$^1$ Kirensky Institute of Physics, Federal Research Center KSC SB RAS, 660036
Krasnoyarsk, Russia\\
$^2$Siberian State Aerospace University, Krasnoyarsk 660014, Russia}
\date{\today}
\begin{abstract}
We report bound states in the radiation continuum (BSCs)  in a
single infinitely long dielectric rod with periodically stepwise
modulated permittivity alternating from $\epsilon_1$ to
$\epsilon_2$. For $\epsilon_2=1$ in air the rod is equivalent to a
stack of dielectric discs with permittivity $\epsilon_1$. Because
of rotational and translational symmetries the BSCs
 are classified by orbital angular momentum $m$ and the Bloch wave vector $\beta$ directed
along the rod. For $m=0$ and $\beta=0$ the symmetry protected BSCs
have definite polarization and occur in a wide range of the radius
of the rod and the dielectric permittivities. More involved BSCs
with $m\neq 0, \beta=0$  exist only for a selected radius of the
rod at a fixed dielectric constant. The existence of robust Bloch
BSCs with $\beta\neq 0, m=0$ is demonstrated. Asymptotic limits to
a homogeneous rod and to very thin discs are also considered.
\end{abstract}
\pacs{42.25.Fx,41.20.Jb,42.79.Dj}
 \maketitle
\section{Introduction}
Recently confined electromagnetic modes above the light line,
bound states in the continuum (BSCs) were shown to exist in (i)
periodic arrays of  long dielectric rods
\cite{Bonnet,Shipman0,Shipman,Marinica,Ndangali2010,Hsu
Nature,Weimann, Wei,Bo
Zhen,Yang,PRA2014,Foley,Hu&Lu,Song,Zou,Yuan,Z Wang,Yuan&Lu,Sadrieva}, (ii)
photonic crystal slabs \cite{Bo Zhen0,Y
Wang,Magnusson,Gao,Blanchard}, and (iii) two-dimensional
periodical structures \cite{Zhang,Kante,Li&Yin} on the surface of
material. Among these different systems the one-dimensional array
of spheres is unique because of rotational symmetry that gives
rise to the BSCs with orbital angular momentum (OAM) \cite{PRA92}.
That reflects in anomalous scattering of plane waves by the array
resulting in scattered electromagnetic fields with OAM travelling
along the array \cite{PRA94,OL,Appl Science,JOSA A}. However,
fabrication of an array of at least hundred identical spheres is a
complicated problem because of technological fluctuations of the
shape of spheres \cite{Adv EM,Peng}. Moreover there is no much
room for tuning parameters of the spheres to achieve BSCs. The
radius can not exceed the half of the period of the array and the
permittivity of the spheres has to be rather high \cite{PRA92}. In
the present paper we consider a single dielectric rod with
periodically modulated permittivity along the rod axis
$\epsilon(z)=\epsilon(z+lh), l=0, \pm 1, \pm2, \ldots$.
\begin{figure}[ht]
\includegraphics[width=10cm,clip=]{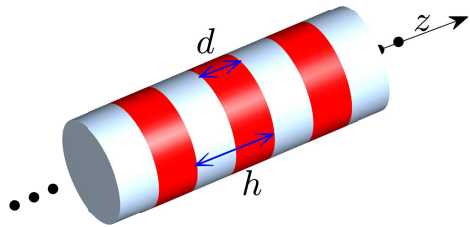}
\caption{(Color online) Infinite circular dielectric rod with
periodically alternating permittivity $\epsilon_1$ (dark red) and
$\epsilon_2$ (light gray).} \label{fig1}
\end{figure}
As shown in Fig. \ref{fig1} for the stepwise behavior with
$\epsilon_2=1$  the rod is equivalent to an one-dimensional array
of dielectric discs with permittivity $\epsilon_1$.
Irrespectively, the rod with periodically modulated permittivity
preserves rotational symmetry. Each dielectric disk has two
geometrical parameters, the radius $R$ and thickness $d$. That
expands the domain of existence of the BSCs to substantially lower
permittivities compared to the case of dielectric spheres.
\section{Eigenmodes with OAM $m=0$}
In what follows we measure all length quantities in terms of the
period $h$ of the array. Because of rotational symmetry the
solutions are classified by integer $m=0, \pm 1, \pm 2, \ldots$,
OAM. At first, we consider TM modes with $m=0$ and $H_r=0, H_z=0,
E_{\phi}=0$ in cylindrical system of coordinates. For that case
our consideration completely follows the approach by Li and
Engheta for plasmonic nanowire \cite{Engheta}. The solution is
sought  in two domains : $r<R$ and $r>R$ independently, and then
matched by the continuity at the rod's boundary $r=R$. We
introduce
\begin{eqnarray}\label{EMTM}
&H_{\phi}(z,r)=\epsilon(z)^{1/2}\psi_{_{TM}}(z,r),&\nonumber\\
&E_r=-\frac{i}{k_0\epsilon(z)}\frac{\partial
\epsilon(z)^{1/2}\psi_{_{TM}}}{\partial z},&\\
&E_z=\frac{i}{k_0\epsilon(z)r}\frac{\partial
\epsilon(z)^{1/2}r\psi_{_{TM}}}{\partial r},&\nonumber
\end{eqnarray}
where the function $\psi_{_{TM}}$ obeys equation
\begin{equation}\label{TMrz}
    \left[\frac{\partial^2}{\partial r^2}+\frac{1}{r}\frac{\partial }{\partial r}
    -\frac{1}{r^2}+\frac{\partial^2}{\partial z^2}+U_{_{TM}}(z)\right]\psi_{_{TM}}(z,r)=0,
\end{equation}
where
\begin{equation}\label{UTM}
    U_{_{TM}}(z)=\epsilon(z)k_0^2-\frac{3}{4}\left(\frac{\epsilon'(z)}
    {\epsilon(z)}\right)^2
    +\frac{1}{2}\frac{\epsilon''(z)}{\epsilon(z)}.
\end{equation}

 For the TE mode in sector $m=0$ we have $E_r=E_z=H_{\phi}=0$ and
\begin{equation}\label{Hr Hz}
E_{\phi}=\psi_{_{TE}}, ~~ H_r=\frac{i}{k_0}\frac{\partial
\psi_{_{TE}}}{\partial z}, ~~ H_z=-\frac{i}{k_0r}\frac{\partial
r\psi_{_{TE}}}{\partial r}
\end{equation}
where the equation for $\psi_{TE}$ has the same form as Eq.
(\ref{TMrz}) except that the effective potential $U_{TM}$ is now
replaced by
\begin{equation}\label{UTE}
    U_{TE}(z)=\epsilon(z)k_0^2.
\end{equation}
Hence we can generalize Eq. (\ref{TMrz}) for both EM modes as
follows
\begin{equation}\label{SE}
    \left[\frac{\partial^2}{\partial r^2}+\frac{1}{r}\frac{\partial }{\partial r}
    -\frac{1}{r^2}+\frac{\partial^2}{\partial z^2}+U_{\sigma}(z)\right]\psi_{\sigma}(z,r)=0,
\end{equation}
where
\begin{equation}\label{psisigma}
    \psi_{\sigma}=\left\{\begin{array}{cc}
E_{\phi}, & \sigma=TE \\ H_{\phi}/\epsilon^{1/2}, & \sigma=TM,
\end{array}\right.
\end{equation}

Because of periodicity of the permittivity the effective potential
$U_{\sigma}(z)$ and the solution of Eq. (\ref{SE}) can be expanded
in Bloch series as
\begin{eqnarray}\label{psiBloch}
&U_{\sigma}(z)=k_0^2\sum_nU_n^{\sigma}e^{iqnz}, q=2\pi/h,&\nonumber\\
&~  \psi_{\sigma}(z,r)=\sum_n\psi_{n\sigma}(r)e^{i(qn+\beta)z},&
\end{eqnarray}
where $\beta$ is the Bloch vector. Then substitution of these
series into Eq. (\ref{SE}) gives
\begin{equation}\label{SEsigma}
    \left[\frac{\partial^2}{\partial r^2}+\frac{1}{r}\frac{\partial }{\partial r}
    -\frac{1}{r^2}-(qn+\beta)^2\right]\psi_{n\sigma}+k_0^2\sum_{n'}U_{n-n'}^{\sigma}\psi_{n'\sigma}=0.
\end{equation}
The presenting the solution as \cite{Engheta}
\begin{equation}\label{psi}
    \psi_{\sigma}(r,z)=\sum_{sn}g_{s\sigma}c_{sn\sigma}J_1(\lambda_{s\sigma}r)
    e^{i(qn+\beta)z},
\end{equation}
we rewrite Eq. (\ref{SEsigma}) in the following form
\begin{equation}\label{cn}
    [-\lambda^2_{\sigma}-(qn+\beta)^2]c_{n\sigma}+k_0^2\sum_{n'}U_{n-n'}^{\sigma}c_{n'\sigma}=0.
\end{equation}
where the eigenvalues $\lambda_{\sigma}$ and eigenvectors
$\mathbf{c}_{\sigma}$ are found from the eigenvalue problem
\begin{equation}\label{eigL}
    \widehat{L}^{\sigma}\mathbf{c}_{\sigma}=\lambda_{\sigma}^2\mathbf{c}_{\sigma}
\end{equation}
with the matrix
\begin{equation}\label{L}
    L_{nn'}^{\sigma}=-(qn+\beta)^2\delta_{nn'}+k_0^2U_{n-n'}^{\sigma}.
\end{equation}

Owing to the equality
$\frac{d}{dx}J_1(x)=J_0(x)-\frac{1}{x}J_1(x)$ we have from Eq.
(\ref{EMTM}) for the TM electric field inside the rod
\begin{equation}\label{Ez}
E_z=\frac{i}{k_0\sqrt{\epsilon}}\sum_{sn}\lambda_{_{s,TM}}~g_{_{s,TM}}~c_{_{sn,TM}}
J_0(\lambda_{_{s,TM}}r)
 e^{i(qn+\beta)z}.
\end{equation}
By the use of the following series
\begin{equation}\label{sqrte}
    \sqrt{\epsilon}=\sum_na_ne^{iqnz}, ~~\frac{1}{\sqrt{\epsilon}}=\sum_nb_ne^{iqnz}
\end{equation}
we obtain  for the components of EM fields at $r\leq R$
\begin{eqnarray}\label{EM inside}
&H_{\phi}=\sum_{snl}g_{_{s,TM}}c_{_{sl,TM}}J_1(\lambda_{_{s,TM}}r)a_{n-l}e^{i(qn+\beta)z}&\nonumber\\
&E_z=\frac{i}{k_0}\sum_{snl}\lambda_{_{s,TM}}g_{_{s,TM}}c_{_{sl,TM}}J_0(\lambda_{_{s,TM}}r)b_{n-l}e^{i(qn+\beta)z}.&
\end{eqnarray}
Outside the rod we have
\begin{eqnarray}\label{EM outside}
&H_{\phi}=\sum_n h_nH_1^{(1)}(\alpha_nr)e^{i(qn+\beta)z},&\nonumber\\
&E_z=\frac{i}{k_0}\sum_n\alpha_nh_nH_0^{(1)}(\alpha_nr)e^{i(qn+\beta)z},&
\end{eqnarray}
where
\begin{equation}\label{alphan}
    \alpha_n=\sqrt{k_0^2-(\beta+qn)^2}
\end{equation}
and $H_1^{(1)}$ and $H_0^{(1)}$ are the Hankel functions. Sewing
at the boundary $r=R$ gives the following dispersion relation
\cite{Engheta}
\begin{equation}\label{dispTM}
    Det(\widehat{S}\widehat{U}\widehat{B}-\widehat{D}\widehat{V}\widehat{T})=0
\end{equation}
where the matrix elements
\begin{eqnarray}\label{SUDT}
&S_{nn'}=\alpha_nH_n^{(0)}(\alpha_nR)\delta_{nn'},&\nonumber\\
&U_{nn'}=a_{n-m}, B_{nn'}=c_{_{nn',TM}}J_1(\lambda{_{n,TM}}R),&\nonumber\\
&D_{nn'}=H_n^{(1)}(\alpha_nR)\delta_{nn'}, V_{nn'}=b_{n-m},&\nonumber\\
&T_{nn'}=c_{_{nn',TM}}\lambda{_{n,TM}}J_0(\lambda{_{n,TM}}R).&
\end{eqnarray}

Respectively for the TE modes we have
\begin{eqnarray}\label{EM inside}
&E_{\phi}=\sum_{sn}g_{_{s,TE}}c_{_{sn,TE}}J_1(\lambda_{_{n,TE}}r)e^{i(qn+\beta)z}&\nonumber\\
&H_z=-\frac{i}{k_0}\sum_{sn}\lambda_{_{s,TE}}g_{_{s,TE}}c_{_{sn,TE}}J_0(\lambda_{_{s,TE}}r)e^{i(qn+\beta)z}.&
\end{eqnarray}
Outside the rod we have
\begin{eqnarray}\label{EM outside}
&E_{\phi}=\sum_nh_nH_1^{(1)}(\alpha_nr)e^{i(qn+\beta)z}&\nonumber\\
&H_z=-\frac{i}{k_0}\sum_n\alpha_nh_nH_0^{(1)}(\alpha_nr)e^{i(qn+\beta)z},&
\end{eqnarray}
Repeating the above algebra  for the TE mode we obtain instead of
(\ref{dispTM}) the following dispersion equation
\begin{equation}\label{dispTE}
    Det(\widehat{S}\widehat{B}-\widehat{D}\widehat{T})=0
\end{equation}
where the matrix elements for all quantities have the form given
by Eq. (\ref{SUDT}) with replacement $TM\rightarrow TE$.
\section{Sectors $m\neq 0$}
Similar to the rod with the homogeneous permittivity for sectors
with $m\neq 0$ the TE and TM solutions are hybridized by the
boundary conditions. Let us start with pure TE mode which can be
expressed through the auxiliary function $\psi_{_{TE}}$:
\begin{eqnarray}\label{function}
&    E_{\phi}=\frac{i}{m}\frac{\partial \psi_{TE}}{\partial r}, E_r=\frac{\psi_{TE}}{r},&\nonumber\\
&    H_{\phi}=-\frac{i}{k_0r}\frac{\partial \psi_{TE}}{\partial z},&\nonumber\\
&H_r=-\frac{1}{k_0m}\frac{\partial^2 \psi_{TE}}{\partial z\partial r},&\nonumber\\
&H_z=\frac{1}{k_0m}\left[\frac{\partial^2}{\partial r^2}
+\frac{1}{r}\frac{\partial}{\partial
r}-\frac{m^2}{r^2}\right]\psi_{_{TE}},&
\end{eqnarray}
 Similarly for the TM mode  we have the following
\begin{eqnarray}\label{function}
&    H_{\phi}=\frac{i\sqrt{\epsilon}}{m}\frac{\partial
\psi_{_{TM}}}{\partial r},~~
H_r=\frac{\sqrt{\epsilon}\psi_{_{TM}}}{r},&\nonumber\\
&   E_{\phi}=\frac{i}{k_0\epsilon r}\frac{\partial \sqrt{\epsilon}\psi_{_{TM}}}{\partial z}, &\nonumber\\
&E_r=\frac{1}{k_0\epsilon m}\frac{\partial \sqrt{\epsilon}}{\partial z}\frac{\partial \psi_{_{TM}}}{\partial r}, &\\
&E_z=-\frac{1}{k_0\sqrt{\epsilon}m}\left[\frac{\partial^2}{\partial
r^2}+ \frac{1}{r}\frac{\partial}{\partial
r}-\frac{m^2}{r^2}\right]\psi_{_{TM}},&\nonumber
\end{eqnarray}
where the auxiliary functions obey the equation
\begin{equation}\label{SEm}
    \left[\frac{\partial^2}{\partial r^2}+\frac{1}{r}\frac{\partial }{\partial r}
    -\frac{m^2}{r^2}+\frac{\partial^2}{\partial z^2}+U_{\sigma}(z)\right]\psi_{\sigma}(z,r)=0.
\end{equation}

The series (\ref{psi}) are modified as follows for both types of
the modes
\begin{equation}\label{psim}
    \psi_{\sigma}(r,z)=\sum_{sn}g_{s,\sigma}c_{sn\sigma}
    J_m(\lambda_{s,\sigma}r)e^{i(qn+\beta)z}.
\end{equation}
Note, the eigenvalues $\lambda_{s,\sigma}$ and eigenvector
amplitudes $c_{sn\sigma}$ coincide with those introduced in the
previous section for $m=0$.  Substituting (\ref{psim}) into Eq.
(\ref{SEm}) and satisfying the boundary conditions, after
cumbersome algebra we obtain the following dispersion relation
\begin{eqnarray}\label{XY}
&im(\widehat{A}-i\widehat{B}-\widehat{I}\widehat{D})\overrightarrow{\psi}_{_{TM}}+
k_0R(\widehat{F}-i\widehat{J}\widehat{P})\overrightarrow{\psi}_{_{TE}}=0,&\nonumber\\
&k_0R(\widehat{K}-i\widehat{J}\widehat{P})\overrightarrow{\psi}_{_{TM}}-
im\widehat{I}(\widehat{C}-\widehat{D})\overrightarrow{\psi}_{_{TE}}=0.&
\end{eqnarray}
where according to Eq. (\ref{psim}) the s-th component of the
vectors $\overrightarrow{\psi}_{\sigma}$ is given by
 \begin{equation}\label{psi vector}
(\overrightarrow{\psi}_{\sigma})_s=g_{s,\sigma}J_m(\lambda_{s,\sigma}R).
\end{equation}
The elements of matrices in Eq. (\ref{XY}) could be found as
\begin{eqnarray}
&A_{ns}=\sum_lb _{n-l}(\beta+ql)c_{sl,_{TM}},&\nonumber\\
&B_{ns}=\sum_ld_{n-l}c_{sl,_{TM}},&\nonumber\\
&I_{nm}=\delta_{nm}(\beta+qn),&\nonumber\\
&J_{nm}=\delta_{nm}\alpha_n\frac{H_m'^{(1)}(\alpha_nR)}{H_m^{(1)}(\alpha_nR)},&\nonumber\\
&F_{ns}=\lambda_{s,_{TE}}c_{sl,_{TE}}\frac{J_m'(\lambda_{s,_{TE}}R)}{J_m(\lambda_{s,_{TE}}R)},&\nonumber\\
&K_{ns}=\lambda_{s_{TE}}\frac{J_m'(\lambda_{s_{TE}}R)}{J_m(\lambda_{s_{TE}}R)}
\sum_la_{n-l}c_{sl,_{TE}},&\nonumber\\
&P_{ns}=\frac{\lambda_{s_{TE}}^2}{\alpha_n^2}c_{ns,_{TE}}.&\nonumber\\
&D_{ns}=\frac{\lambda_s^2}{\alpha_n^2}\sum_lb_{n-l}c_{sl,_{TM}},&
\end{eqnarray}
where
\begin{equation}\label{D}
    \frac{\epsilon'(z)}{2\epsilon^{3/2}(z)}=\sum_nd_{n}e^{iqnz}.
\end{equation}

In order to avoid discontinuities of the derivatives of the
permittivity at the boundary of the disc $z=\pm 1/2$ we following
Ref. \cite{Engheta} smooth the boundary by the function $$
\epsilon(z)=\epsilon_2+\frac{1}{2}(\epsilon_1-\epsilon_2)[1-\tanh(\kappa
(|z|-1/2))]$$  with the control parameter $\kappa$. In what
follows we take $\kappa=17$.

\section{Symmetry classification of BSCs}
Similar to the periodic array of dielectric spheres the BSCs in
the single rod with periodically modulated permittivity are
classified by the OAM $m$ due to the rotational symmetry of the
rod and the Bloch vector along the rod due to the translational
symmetry. Moreover there is the mirror symmetry $z\rightarrow -z$.
That allows us to classify the BSCs with $\beta=0$ by parity.
These standing wave BSCs are symmetry protected relative to ever
the TE diffraction continuum or the TM continuum. Introduce the
operator $\widehat{O}f(z)=f(-z)$. Respectively after the Fourier
transformation we have $\widehat{O}f_n=f_{-n}$ and therefore
$O_{nn'}=\delta_{n+n',0}$. The operator $\widehat{L}^{\sigma}$
with matrix elements given by Eq. (\ref{L}) for $\beta=0$ commutes
with the operator $\widehat{O}$. Therefore the eigenvectors of the
operator $\widehat{L}^{\sigma}$ are classified as even and odd
\begin{equation}\label{mir}
    c_{sn,\sigma}=\pm c_{s,-n,\sigma}.
\end{equation}

Let us rewrite Eq. (\ref{XY}) as follows
\begin{eqnarray}\label{HH}
&\widehat{H}_{_{1,TM}}\overrightarrow{\psi}_{_{TM}}+
\widehat{H}_{_{1,TE}}\overrightarrow{\psi}_{_{TE}}=0,&\nonumber\\
&\widehat{H}_{_{2,TM}}\overrightarrow{\psi}_{_{TM}}+\widehat{H}_{_{2,TE}}
\overrightarrow{\psi}_{_{TE}}=0,&
\end{eqnarray}
where matrices $\widehat{H}_{k,\sigma}, k=1,2$ are of the size
$(2N+1)\times (2N+1)$. We arrange the matrices as follows
\begin{eqnarray}\label{HHH}
&\widehat{H}_{_{1,TE}}=[\widehat{H}_{_{1,TE}}^e\{(2N+1)\times(N+1)\},
~~ \widehat{H}_{_{1,TE}}^o\{(2N+1)\times N\}],&\nonumber\\
&\widehat{H}_{_{1,TM}}=[\widehat{H}_{_{1,TM}}^e\{(2N+1)\times N\},
~~ \widehat{H}_{_{1,TM}}^o\{(2N+1)\times(N+1)\}],&\nonumber\\
&\widehat{H}_{_{2,TE}}=[\widehat{H}_{_{2,TE}}^o\{(2N+1)\times(N+1)\},
~~ \widehat{H}_{_{2,TE}}^e\{(2N+1)\times N\}],&\\
&\widehat{H}_{_{2,TM}}=[\widehat{H}_{_{2,TM}}^o\{(2N+1)\times N\},
~~ \widehat{H}_{_{2,TM}}^e\{(2N+1)\times (N+1)\}]&\nonumber
\end{eqnarray}
 where
expressions in curly brackets show the size of the matrices and
the matrix elements are even or odd relative to $n\rightarrow -n$:
\begin{equation}\label{HeHo}
\widehat{H}_{nn',\sigma}^e=\widehat{H}_{-nn',\sigma}^e, ~~
\widehat{H}_{nn',\sigma}^o=-\widehat{H}_{-nn',\sigma}^o.
\end{equation}

Substituting relations (\ref{HHH}) into Eq. (\ref{HH}) and
splitting the vector
$$\overrightarrow{\psi}_{_{TE}}=\left(\begin{array}{c}
  \overrightarrow{\psi}_{\uparrow_{TE}}\{N+1\} \\
  \overrightarrow{\psi}_{\downarrow_{TE}}\{N\} \\
  \end{array}\right), ~~
  \overrightarrow{\psi}_{_{TM}}=\left(\begin{array}{c}
  \overrightarrow{\psi}_{\uparrow_{TM}}\{N\} \\
  \overrightarrow{\psi}_{\downarrow_{TM}}\{N+1\} \\
  \end{array}\right),$$
we obtain the following equations
\begin{eqnarray}\label{eo}
&\widehat{H}_{_{1,TM}}^e\overrightarrow{\psi}_{\uparrow_{TM}}+
\widehat{H}_{_{1,TM}}^o\overrightarrow{\psi}_{\downarrow_{TM}}+
\widehat{H}_{_{1,TE}}^e\overrightarrow{\psi}_{\uparrow_{TE}}+
\widehat{H}_{_{1,TE}}^o\overrightarrow{\psi}_{\downarrow_{TE}}=0,&\nonumber\\
&\widehat{H}_{_{2,TM}}^o\overrightarrow{\psi}_{\uparrow_{TM}}+
\widehat{H}_{_{2,TM}}^e\overrightarrow{\psi}_{\downarrow_{TM}}+
\widehat{H}_{_{2,TE}}^o\overrightarrow{\psi}_{\uparrow_{TE}}+
\widehat{H}_{_{2,TE}}^e\overrightarrow{\psi}_{\downarrow_{TE}}=0.&
\end{eqnarray}
From Eqs. (\ref{HeHo}) and (\ref{eo}) it follows that there are
two solutions. The first is $\overrightarrow{\psi}_{\downarrow
\sigma}= 0$ and $\overrightarrow{\psi}_{\uparrow \sigma}\neq 0$
with $H_z, E_{\phi}$, and $E_r$ even  and $E_z, H_{\phi}$ and
$H_r$ odd relative to the inversion $z\rightarrow -z$. This
solution gives us a TM symmetry protected BSC. The second solution
$\overrightarrow{\psi}_{\uparrow \sigma}= 0$ and
$\overrightarrow{\psi}_{\downarrow \sigma}\neq 0$ has odd field
components $H_z, E_{\phi}$, and $E_r$ even $E_z, H_{\phi}$ and
$H_r$. This solution is a TE symmetry protected BSC. By solving
Eqs. (\ref{dispTM}), (\ref{dispTE}) and (\ref{XY}) numerically we
obtain the following set of BSCs. In particular there are symmetry
protected BSCs with definite polarization which occur at arbitrary
radius of the rod:

(1) Symmetry protected TE BSCs with $\beta=0, m=0$ and
$H_z(-z)=-H_z(-z)$.

(2) Symmetry protected TM BSCs with $\beta=0, m=0$ and
$E_z(-z)=-E_z(-z)$.

\begin{figure}[ht]
\includegraphics[width=10cm,clip=]{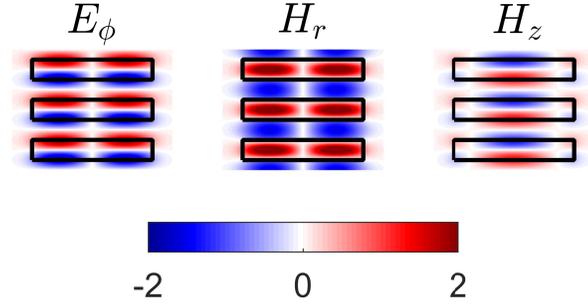}
\caption{Pattern of the symmetry protected  TE BSC  with zero OAM
$m=0$, frequency $k_{0c}= 4.6063$ and $\beta=0$ for parameters:
$\epsilon_1=3, \epsilon_2=1, R=1.5, d=0.5$.} \label{fig2}
\end{figure}

\begin{figure}[ht]
\includegraphics[width=11cm,clip=]{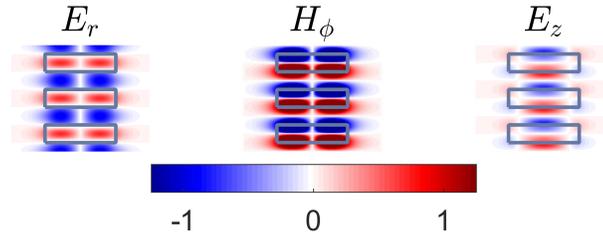}
\caption{Pattern of the symmetry protected  TM BSC  with zero OAM
$m=0$, frequency $k_{0c}= 5.37652$ and $\beta=0$ for parameters:
$\epsilon_1=3, \epsilon_2=1, R=1, d=0.5$.} \label{fig3}
\end{figure}

Examples of these symmetry protected BSCs are shown in Fig.
\ref{fig2} and Fig. \ref{fig3}.

The next class of the BSCs with definite polarization are
non-symmetry protected and require tuning the rod radius $R$:

(3) Non-symmetry protected TE BSCs with $\beta=0, m=0$ and
$H_z(-z)=H_z(-z)$.

(4) Non-symmetry protected TM BSCs with $\beta=0, m=0$ and
$E_z(-z)=E_z(-z)$.

\begin{figure}[ht]
\includegraphics[width=11cm,clip=]{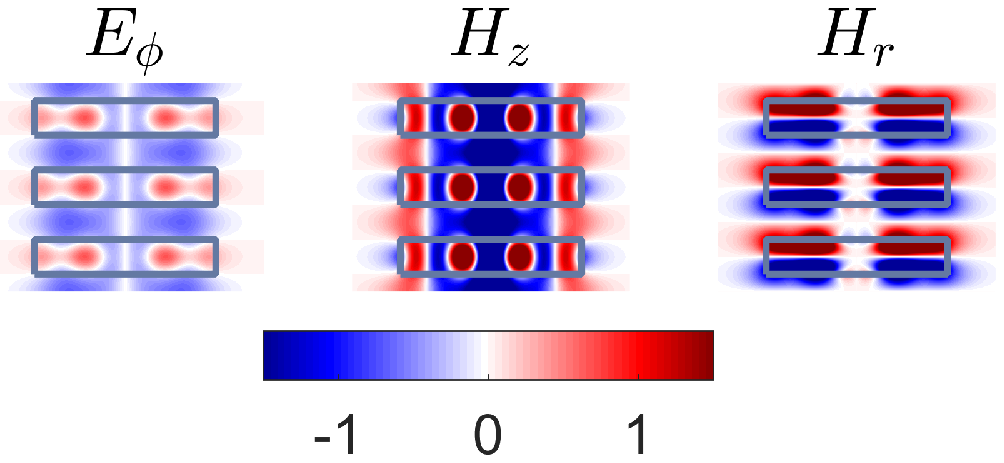}
\caption{Pattern of the non symmetry protected  TE BSC  with $m=0,
\beta=0$ and frequency $k_{0c}= 4.63778$ for parameters:
$\epsilon_1=5, \epsilon_1=1,  R=1.3061, d=0.5$.} \label{fig4}
\end{figure}

\begin{figure}[ht]
\includegraphics[width=11cm,clip=]{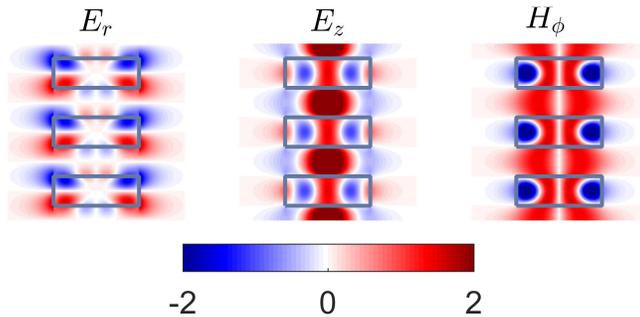}
\caption{Pattern of the non symmetry protected  TM BSC  with $m=0,
\beta=0$ and frequency $k_{0c}= 4.85502$ for parameters:
$\epsilon_1=5, \epsilon_2=1, R=0.724504, d=0.5$.} \label{fig5}
\end{figure}

These BSCs are  shown in Fig. \ref{fig4} and Fig. \ref{fig5}.

\begin{figure}[ht]
\includegraphics[width=11cm,clip=]{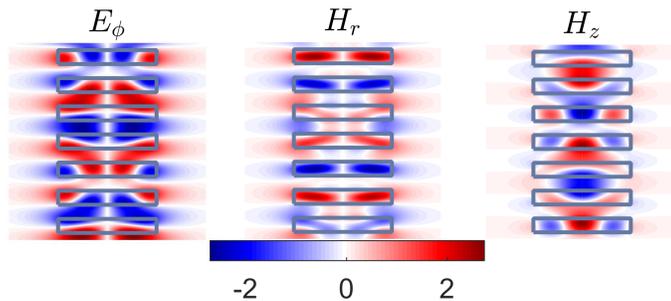}
\caption{(Color online) Pattern of the Bloch  TE BSC  with $m=0,
\beta_c=2.37361$ and frequency $k_{0c}= 3.15725$ for parameters:
$\epsilon_1=5, R=1, d=0.5$.} \label{fig6}
\end{figure}

\begin{figure}[ht]
\includegraphics[width=11cm,clip=]{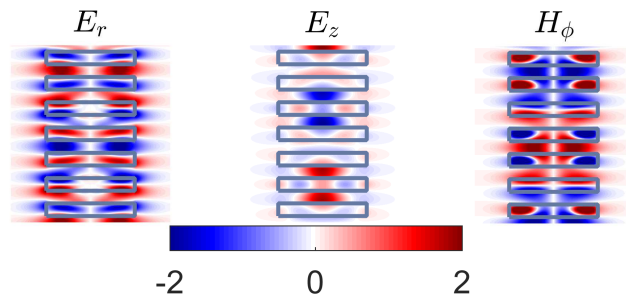}
\caption{(Color online) Pattern of the Bloch  TM BSC  with $m=0,
\beta_c=1.08676$ and frequency $k_{0c}= 3.88423$ for parameters:
$\epsilon_1=5, R=1, d=0.5$.} \label{fig7}
\end{figure}

(5) Bloch BSCs with $\beta\neq 0, m=0$ with definite polarization
shown in Fig. \ref{fig6} and Fig. \ref{fig7}. They exist within a
wide interval of the rod radius. Rigorously speaking the Bloch
BSCs can not be considered as guided modes similar to those  which
exist below light line in the homogeneous dielectric rod
\cite{Jackson}. However those Bloch quasi-BSCs in some small
interval of $\beta$ around the BSC point have the lifetimes
exceeding the propagation time in the rod of finite length and
thus can be considered as the guided modes above the light line
\cite{OL}.

\begin{figure}[ht]
\includegraphics[width=10cm,clip=]{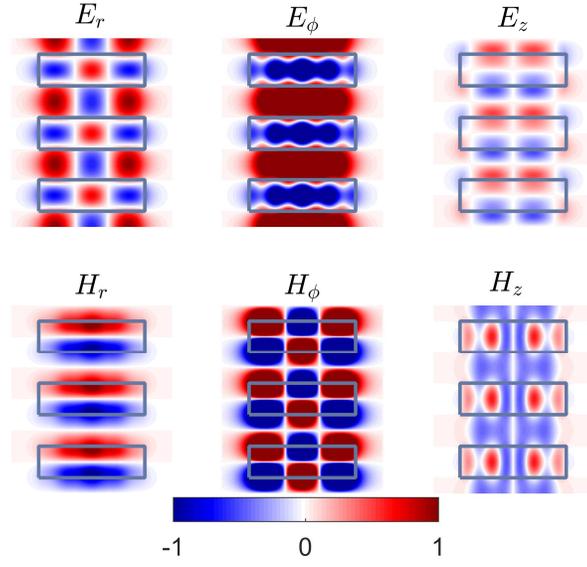}
\caption{(Color online) Pattern of the  BSC  with $m=1$ symmetry
protected in respect to the TM radiation continuum and frequency
$k_{0c}=5.26284$ for tuned radius $R=1.65293$.} \label{fig8}
\end{figure}

\begin{figure}
\includegraphics[width=10cm,clip=]{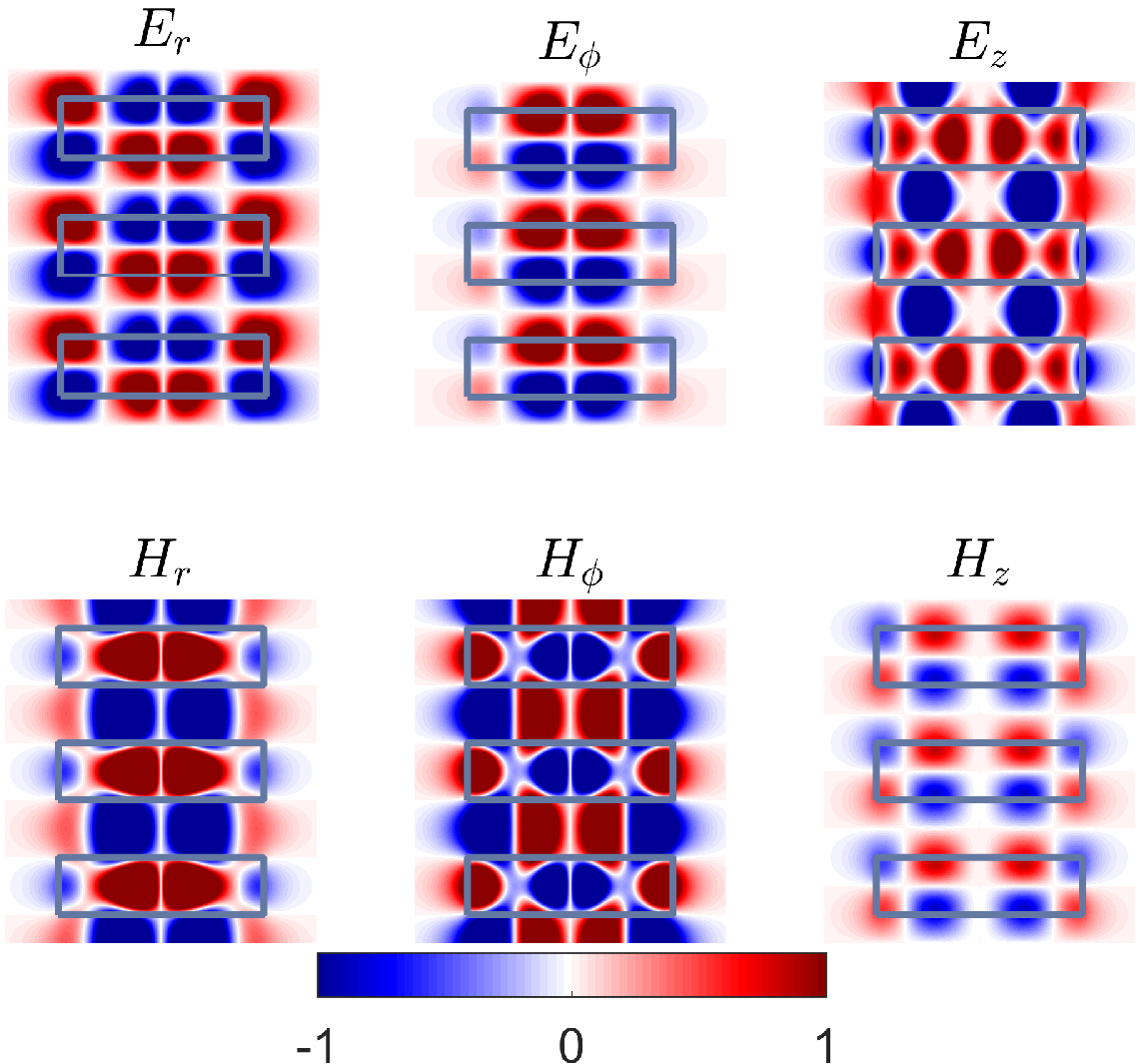}
\caption{(Color online) Pattern of the BSC  with $m=2$ symmetry
protected in respect to the TE radiation continuum and frequency
$k_{0c}=5.21418$ for tuned radius $R=1.7009$.} \label{fig9}
\end{figure}

\begin{figure}[ht]
\includegraphics[width=10cm,clip=]{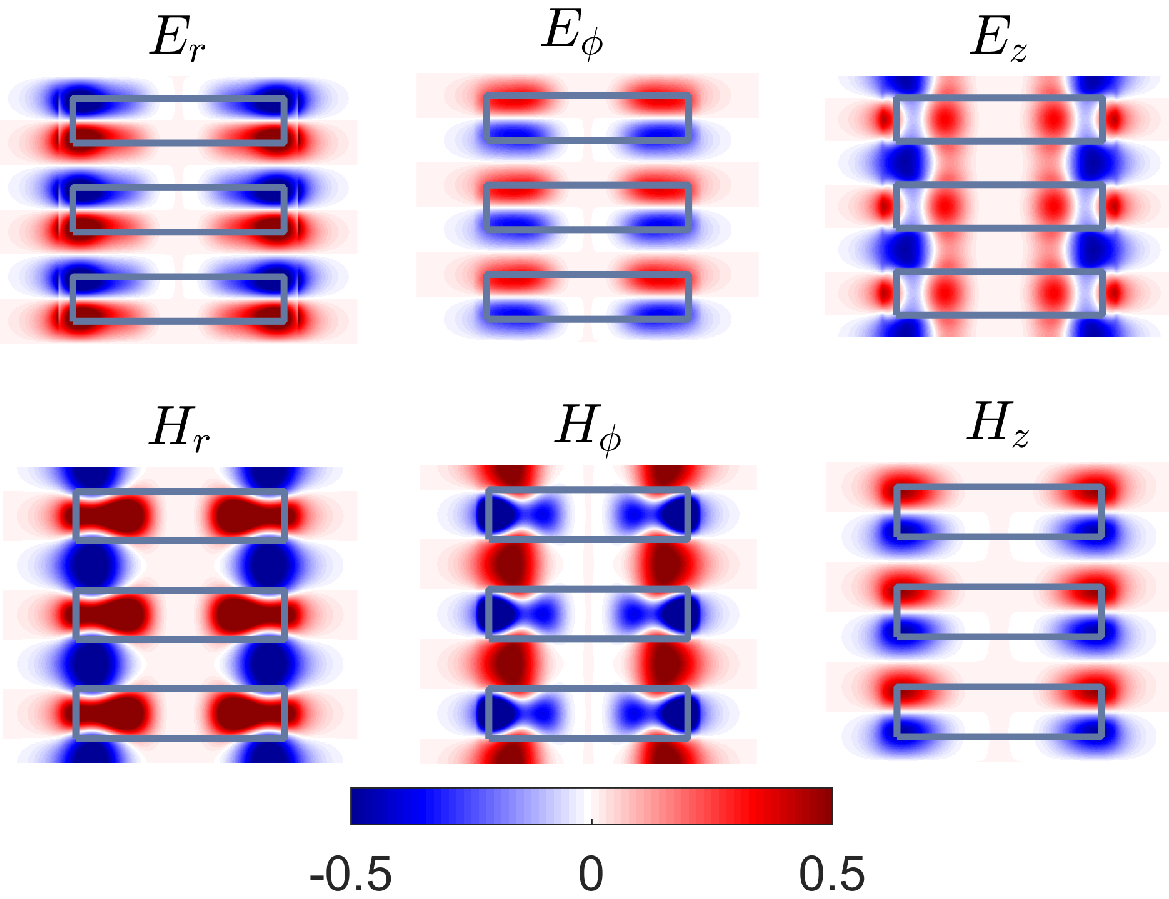}
\caption{(Color online) Pattern of the BSC  with $m=5$ symmetry
protected in respect to the TE radiation continuum and frequency
$k_{0c}=5.14387$ for tuned radius $R=1.87591$.} \label{fig10}
\end{figure}

\begin{figure}[ht]
\includegraphics[width=10cm,clip=]{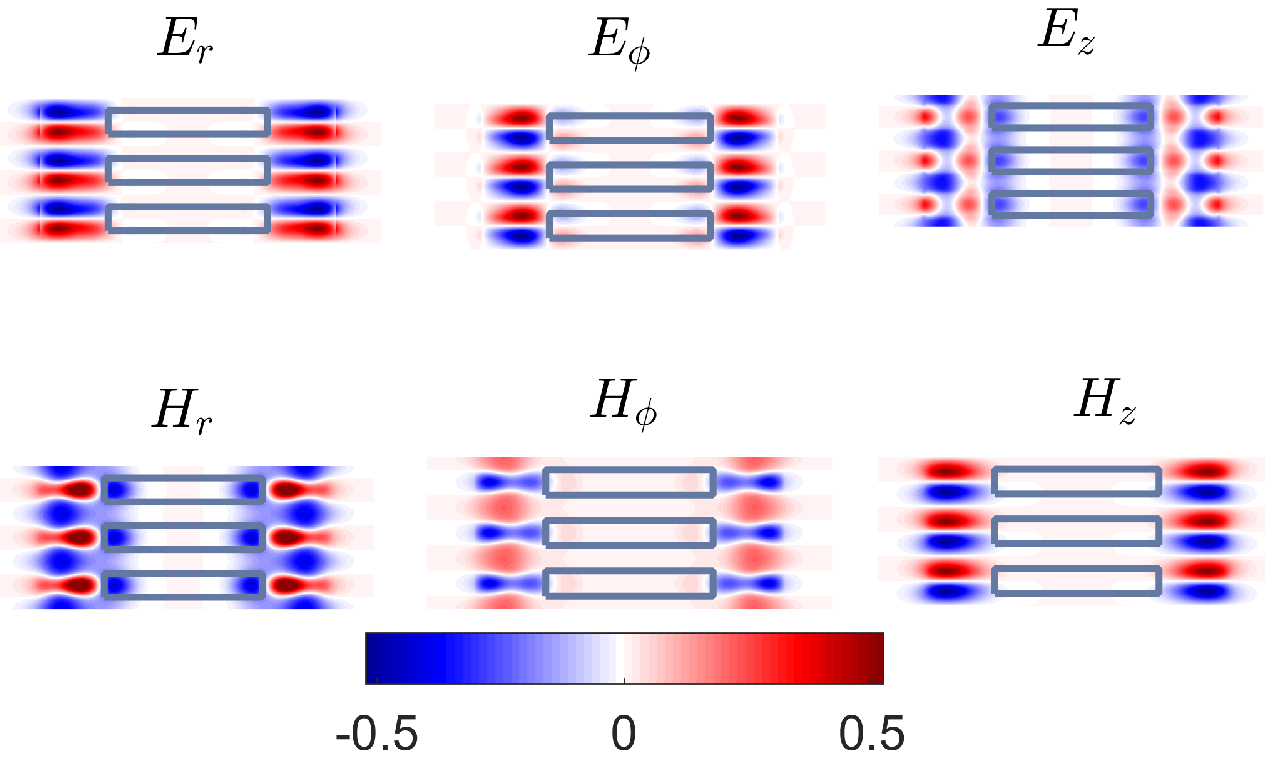}
\caption{(Color online) Pattern of the BSC  with $m=10$ symmetry
protected in respect to the TE radiation continuum and frequency
$k_{0c}=5.29052$ for tuned radius $R=3.07046$.} \label{fig11}
\end{figure}

(6) BSCs with orbital angular momentum (OAM) $m\neq 0$ and
$\beta=0$ constitute the most interesting class. Whilst in the
array of spheres we managed to find only BSCs with $m=1$ and $m=2$
\cite{PRA94,Adv EM}. In the array of discs we found BSCs with
higher OAM. However, in contrast to the array of spheres we did
not find any Bloch BSCs with $m\neq 0$ and $\beta \neq 0$. The
BSCs with OAM are hybridized with respect to polarization. They
are symmetry protected against decay into the TE/TM continuum as
it was considered above but the radius has to be tuned for the
mode be decoupled from the TM/TE continuum. Figs.
\ref{fig8}--\ref{fig11} show the solutions of Eq. (\ref{XY}) for
BSCs with $m=1, 2, 5, 10$, and $\beta=0$. All BSCs with nonzero
OAM were calculated for $\epsilon_1=3, \epsilon_2=1$ and $d=0.5$.
One can see from Figs. \ref{fig10} and \ref{fig11} a tendency of
light localization at the surface of the rod with growth of the
OAM $m$ limiting to whispering gallery modes. However, in contrast
to the latter  the BSCs with OAM exist for any $m$.

\begin{figure}[ht]
\includegraphics[width=10cm,clip=]{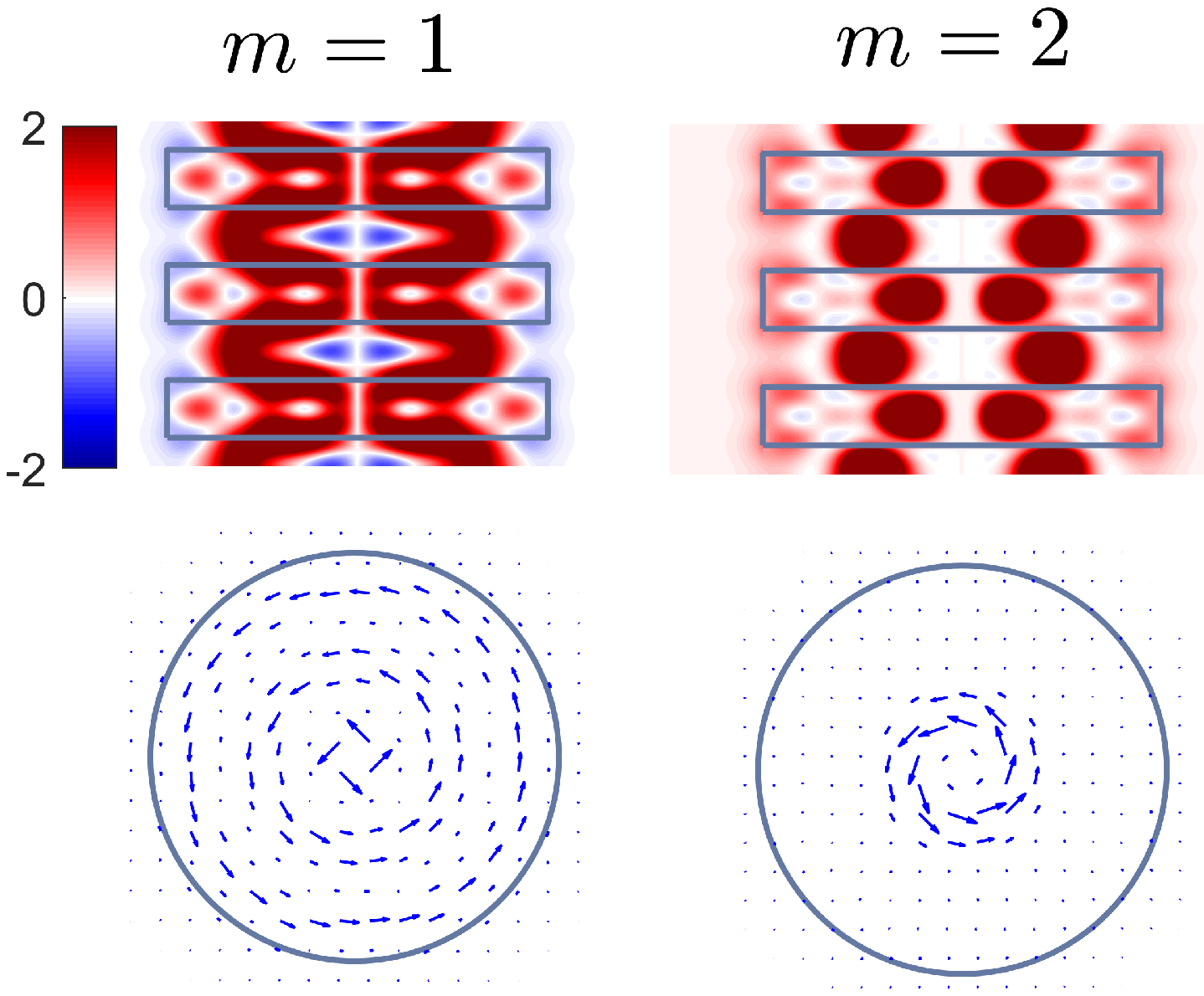}
\includegraphics[width=10cm,clip=]{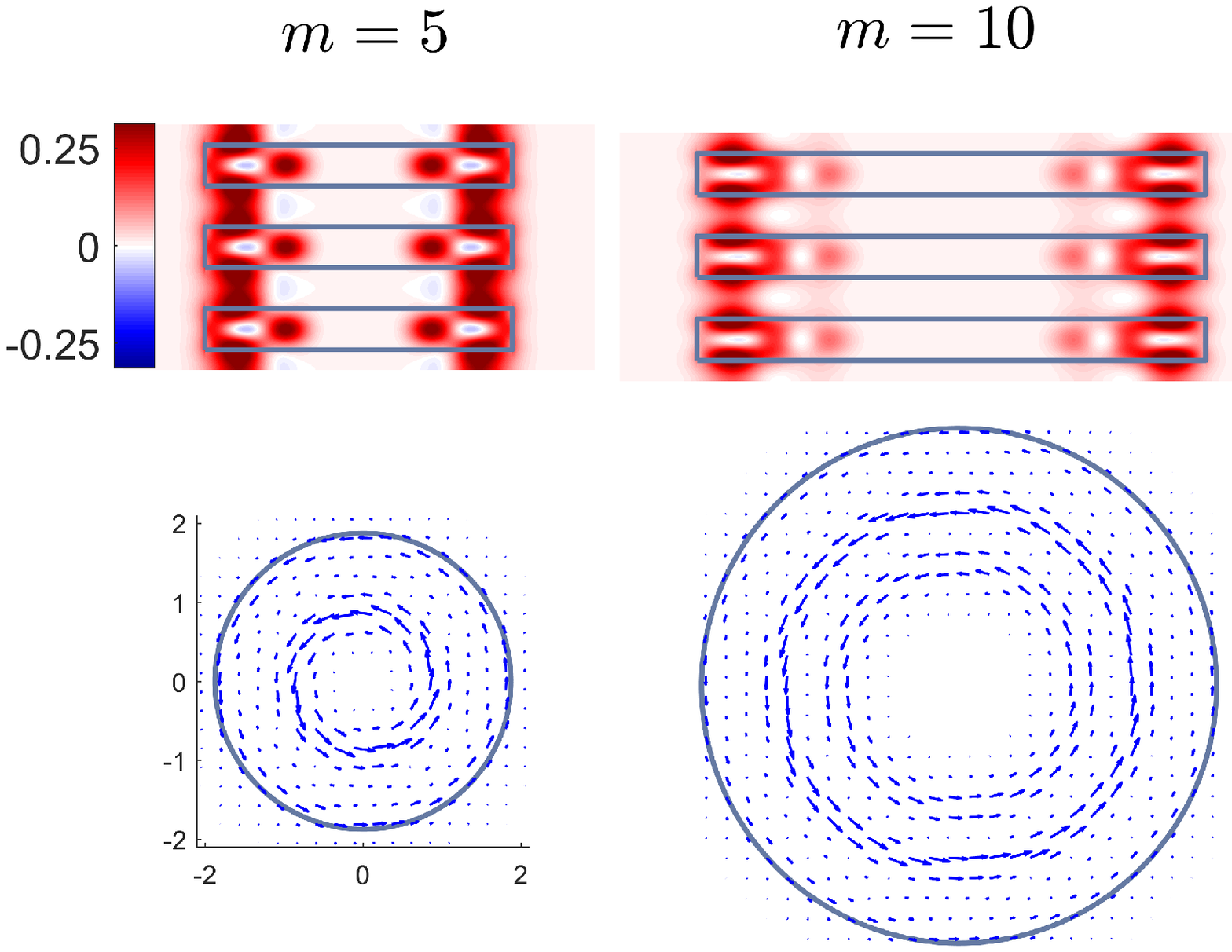}
\caption{(Color online) The profile $H_z$ and flows of Pointing
vector at $z=0$ circulating around core of the rod in the BSCs
shown in Fig. \ref{fig11}.} \label{fig12}
\end{figure}

The BSC with OAM is degenerate with respect to the sign of $m$.
The sign controls the direction of spinning of the Poynting vector
$\overrightarrow{j}=j_0 \overrightarrow{E}\times
\overrightarrow{H}$ as demonstrated in Fig. \ref{fig12}. We
mention in passing that the spinning trapped modes in an acoustic
cylindrical infinitely long waveguide which contains rows of large
numbers of blades arranged around a central core was first
reported by Duan and McIver \cite{Duan}.
\newpage
\section{Limits of the BSCs for $d\rightarrow 1$ and $d \ll 1$}
Until now we considered trapping of light by a stack of
dielectric discs whose thickness equals half of the period. In
this section we consider what happens with the BSC when the rod becomes homogeneous
and when the discs become veru thin.
\begin{figure}[ht]
\includegraphics[width=7cm,clip=]{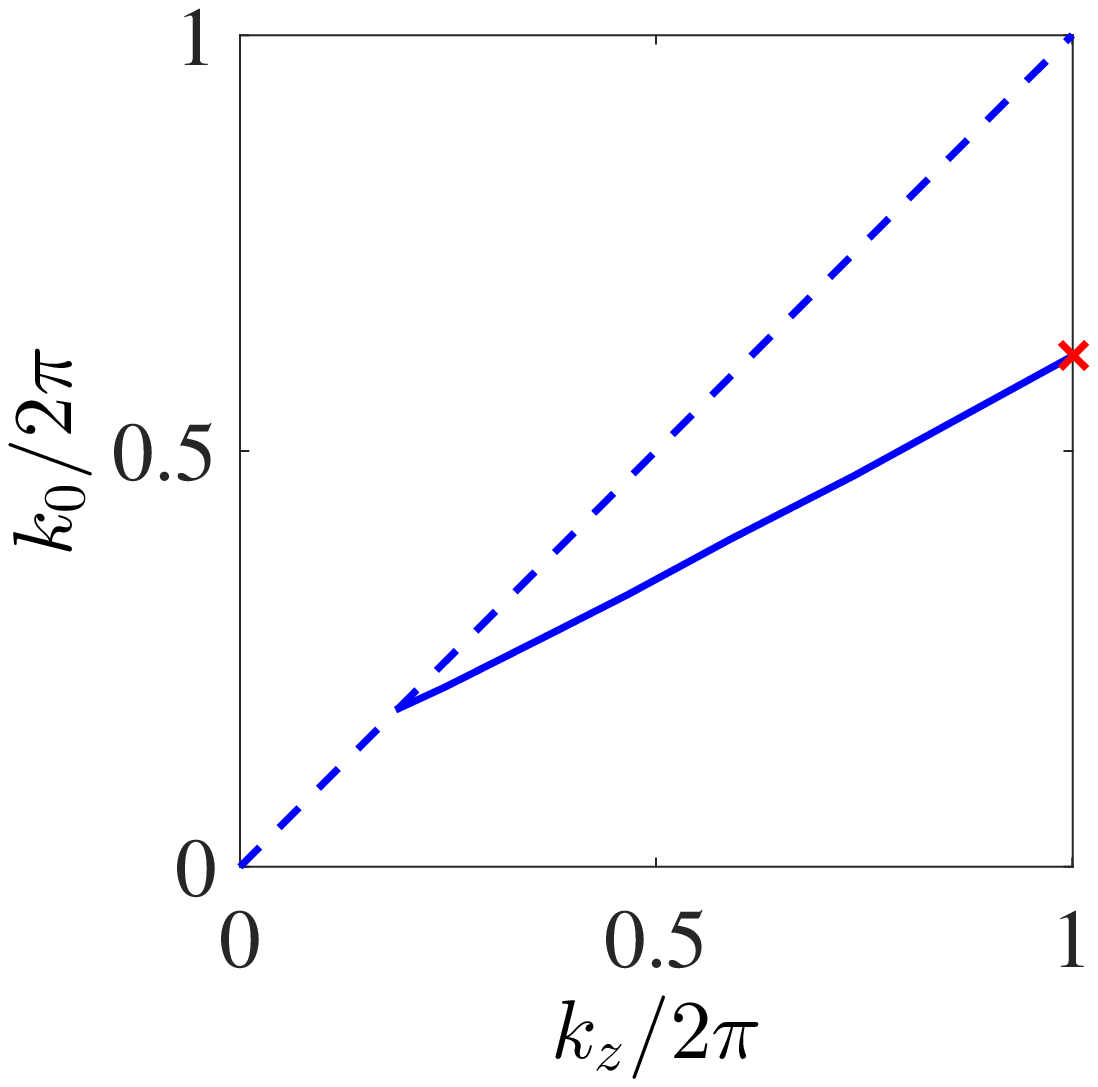}
\caption{Dispersion curve of the waveguide mode, bound state below the light line
in the homogeneous rod with the radius $R=1.5$ and permittivity $\epsilon_1=3$.}
\label{fig14}
\end{figure}
The homogeneous rod which can support
only guided modes with $k_z>0$ below the light line. In the latter the Maxwell
equations can be solved by separation of variables for the TE
polarization with zero OAM $m=0$ \cite{Jackson}
\begin{equation}\label{rod function}
H_z(r,z)=\left\{\begin{array}{cc}
 e^{ik_zz}J_0(\sqrt{\epsilon k_0^2-k_z^2}r)& r\leq R,\\
  Ae^{ik_zz}K_0(\sqrt{k_z^2-k_0^2}r)=\frac{i\pi A}{2}e^{ik_zz}H_0^{(1)}(\sqrt{k_0^2-k_z^2}r)& r>R,
\end{array}\right.
\end{equation}
to result in guided mode, bound state below the light line $k_z <
k_0$ after matching at $r=R$. Numerical result for the dispersion
curve of the lowest TE mode in the homogeneous cylindrical rod is shown in Fig. \ref{fig14} where the
frequency of this solution $k_0=3.858$ at $k_z=2\pi$ is marked by
cross.

\begin{figure}[ht]
\includegraphics[width=7cm,clip=]{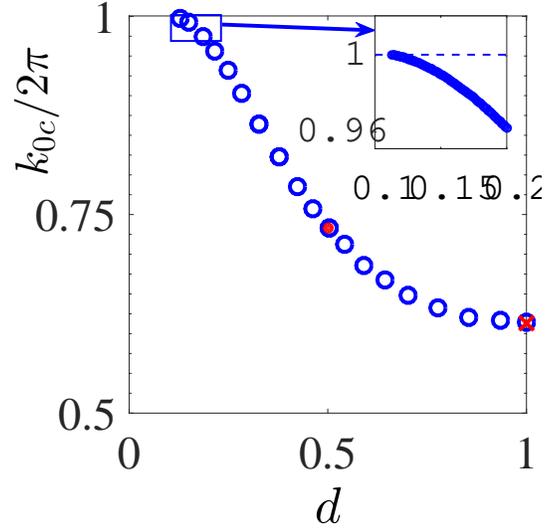}
\caption{Frequency of the TE symmetry protected BSC vs the
thickness of discs in terms of the period $h$ for $R=1.5$. Closed circle notes
the BSC shown in Fig. \ref{fig2} for $d=0.5$.}
\label{fig15}
\end{figure}

As soon as the rod acquires a periodic modulation of the
permittivity $\epsilon(z)=\epsilon(z+l), ~~l=0, \pm 1, \pm
2,\ldots$ the radiation continua in the form of the Hankel
functions (\ref{rod function}) is quantized $k_{z,n}=\beta+2\pi
n$. In the other words, the rod can be viewed as one-dimensional
cylindrical diffraction lattice \cite{Ndangali2010,PRA92}. Let us consider the
TE BSC symmetry protected against the lowest diffraction continuum
$n=0$ above light line shown by dash line in Fig.
\ref{fig14}. Its solution  takes the following form for $r>R$
\begin{equation}\label{psim=0}
    H_z(r,z)=-\frac{i}{k_0}\sum_nh_n\alpha_nH_0^{(1)}(\alpha_nr)\sin(2\pi nz)
\end{equation}
where $a_n$ are given by Eq. (\ref{alphan}).
In particular this solution turns to the symmetry protected TE BSC shown in Fig. \ref{fig2}
if all $h_n=0$ except $h_1$. The dependence of the BSC frequency on the disc thickness is
shown in Fig. \ref{fig15} which limits to the value
$k_{0c}=3.858$ which is just the the frequency of the solution of the homogeneous rod
(\ref{rod function}) marked by cross in Figs. \ref{fig14} and \ref{fig15}.
The solution is very similar to
that shown in Fig. \ref{fig2} but is more localized.

In the second limit when the disk thickness $d$ decreases the mean
permittivity of the rod drops as well and respectively the BSC
frequency grows as plotted in Fig. \ref{fig15}.
\begin{figure}[ht]
\includegraphics[width=10cm,clip=]{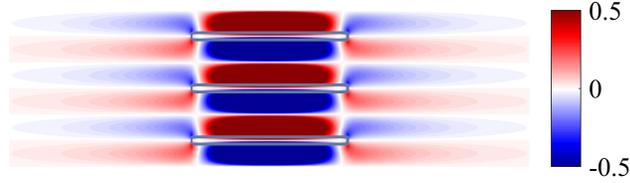}
\caption{Pattern of the TE symmetry protected BSC at $d=0.132,
R=1.5$. } \label{fig16}
\end{figure}
The further decrease of the thickness $d$ brings the BSC frequency
to the bottom of the second diffraction continuum $2\pi$ where the
BSC is corrupted by leakage into that continuum. In the zoomed
window in Fig. \ref{fig15} we show it occurs at $d=0.112$ for
$\epsilon_1=3$. Thus the thickness of disks is limited for the TE
symmetry protected BSC to exist. The radius of localization of the
BSC behaves as
\begin{equation}\label{BSC length}
R_c\approx \frac{1}{\sqrt{4\pi^2-{k_{0c}^2}}}.
\end{equation}
Fig. \ref{fig16} illustrates the $H_z$ component of the BSC
solution near the bottom of the second diffraction continuum at
$d=0.137$.  One can see that the radius of localization is
tremendously increased compared to the case $d=0.5$ shown in Fig.
\ref{fig2}. According to Eq. (\ref{BSC length}) the radius of
localization of the BSC goes to infinity when $d\rightarrow
0.112$.
\section{Summary}
We considered light trapping in a single infinitely long
dielectric rod with periodically  modulated permittivity. We
restrict ourselves with stepwise behavior of the permittivity
intermittently changing from $\epsilon_2=1$ to $\epsilon_1 > 1$ to
makes the rod equivalent to a stack of dielectric discs. Even in
that particular case owing to the possibility of tuning two
dimensional parameters, the radius and thickness of the discs and
the permittivity we have an abundance of BSCs compared to the
array of dielectric spheres \cite{Adv EM}. The stack of discs
preserves the rotational symmetry to give rise to BSCs with
definite OAM. However, in contrast to the array of spheres the rod
with periodically modulated permittivity supports BSCs with OAM up
to $m=10$ for a sufficiently large radius as shown in Fig.
\ref{fig10}. We also found Bloch BSCs with both polarizations
however, only with zero OAM. Bloch BSCs with non zero
OAM have not been found yet.

In the limit $d\rightarrow 1$ when the discs mold into single
homogeneous rod we have shown that the symmetry protected BSC with
$m=0$ transforms into the guided mode below the light line. In the
limit of thin discs $d\ll 1$ the BSC frequency reaches the bottom
of the second diffraction continuum and is destroyed by leakage
into that continuum. The problem of BSCs can be also solved for
sinusoidal behavior $\epsilon(z)=\epsilon_0+\lambda\sin 2\pi z
$.

{\bf Acknowledgments}: We acknowledge discussions with  D.N.
Maksimov and A.S. Aleksandrovsky. This  work  was partially supported  by
Ministry  of  Education  and  Science  of  Russian  Federation
(State contract  N  3.1845.2017) and RFBR grant 16-02-00314.

\end{document}